\documentclass[preprint,12pt]{achemso}
\usepackage[version=3]{mhchem} % Formula subscripts using \ce{}
\usepackage[T1]{fontenc}       % Use modern font encodings
\usepackage[utf8]{inputenc}
\pdfoutput=1

\author{Lionel Chiron}
\affiliation[CASC4DE]{CASC4DE, Le Lodge, 20 Av du neuhof , 67100 Strasbourg, France}

\author{Marie-Aude Coutouly}
\affiliation[NMRTEC]{NMRTEC, Bld. Sébastien Brandt, Bioparc - Bat. B, 67400 Illkirch Graffenstaden, France}
\altaffiliation{DataStorm, 60, rue Etienne Dolet, 92240 Malakoff, France}

\author{Jean-Philippe Starck}
\affiliation[CASC4DE]{CASC4DE, Le Lodge, 20 Av du neuhof , 67100 Strasbourg, France}

\author{Christian Rolando}
\affiliation[MSAP]
{Miniaturisation pour la Synth\`{e}se, l'Analyse et la Prot\'{e}omique (MSAP), USR CNRS 3290, and Prot\'{e}omique, Modifications Post-traductionnelles et Glycobiologie, IFR 147 and Institut Eug\`{e}ne-Michel Chevreul, FR CNRS 2638, Universit\'{e} de Lille 1 Sciences et Technologies, 59655 Villeneuve d'Ascq Cedex, France}

\author{Marc-Andr\'{e} Delsuc}
\affiliation[IGBMC]
{Institut de G\'{e}n\'{e}tique et de Biologie Mol\'{e}culaire et Cellulaire (IGBMC), INSERM, U596; CNRS, UMR 7104, 67404 Illkirch-Graffenstaden, France }
\email{madelsuc@unistra.fr}

\title[SPIKE, a dedicated FT Software]
  {SPIKE a Processing Software dedicated to Fourier Spectroscopies}
\abbreviations{NMR, FT-MS, 2D-FTICR-MS}

\begin{document}


\begin{abstract}
  We present SPIKE (Spectrometry Processing Innovative KErnel), an
  open-source Python package dedicated to Fourier spectroscopies. It
  provides basic functionalities such as apodisation, a complete set of
  Fourier transforms, phasing (for NMR), peak-picking, baseline correction
  and also tools such as Linear Prediction. Beside its versatility, the
  most prominent novelty of this package is to incorporate new tools for
  Big Data processing. This is exemplified by its ability to handle the
  processing and visualization of very large data-sets, with
  multiprocessor capabilities and a low memory footprint. The software
  contains also all the tools necessary for the specific fast processing
  and visualization of 2D-FTICR-MS data-sets.
\end{abstract}

\section{Introduction}

Fourier spectroscopies concept revolutionized many analytical techniques in the second half of the
{$\mathrm{X \! X^{th}}$} century: Nuclear Magnetic Resonance ({NMR}), Infrared spectroscopy ({IR}), {Raman spectroscopy}, and Mass Spectrometry ({MS}) with  Fourier Transform Ion Cyclotronic Resonance (FTICR-MS)\cite{Comisarow1974,Marshall1998} and more recently Orbitrap\cite{Qizhi2005}.
The use of Fast Fourier Transform gives access to Fellgett advantage\cite{fellgett1949} and permits higher sensitivity and faster acquisitions.
As these analytical techniques share the same processing procedures it is possible to devise common tools.
However, generally each domain usually comes with specific methods for display as well as specific file formats, limiting the sharing of tools and leading to redundant developments.

Every experimental study is the result of a three steps procedure.
Data are first acquired, then processed and finally analyzed.
Depending on the domain, processing is a more or less neglected task.
For example, in Mass Spectrometry, analysis is a very important part of the work for the exploitation of the data and many commercial and open-source softwares have been developed for this purpose during the last decades and are now
available\cite{Perkins1999,Geer2004,Sturm2008,Rost2013,Strohalm2008}.
But very few tools are proposed for processing the data before analysis.
Most of the time, this step is abandoned to manufacturers and considered as black
boxes full of obscure procedures, whereas processing is a crucial step for providing a more precise and complete information for analysis.
In contrast, processing is a dominant part of the NMR spectroscopist culture with a very large wealth of existing softwares
\cite{delaglio1995,Pons1996,hoch1996,gunther2000,Tramesel2007,vanbeek2007,helmus2013}.
In addition new processing techniques are profundly modifying data processing and there is a need of incorporating them in the toolbox of the practitioners.

Recent computer capabilities have indeed given the possibility to store very large amount of data at each experimental run.
The rising usage of high-throughput\cite{Nilsson2010} acquisition in Mass Spectrometry has rapidly taken advantage of those progresses.
But in spite of hardware enhancements, the classical processing techniques at disposal (such as denoising, deconvolution, etc..) are facing both the huge volume of data and the limited time available for computations.
This phenomenon, known as the \emph{Big Data bottleneck},
raises the need of newer algorithms, adapted to Big Data by being faster and with less memory footprint.

As a first answer to these two major concerns: generic processing tools for FT spectroscopies and Big Data challenge,
we propose here a framework able to handle general processing and which can currently be used for FT Mass Spectrometry (FTICR and Orbitrap) as well as for NMR processing.
We called it SPIKE (Spectrometry Processing Innovative KErnel), as it is developed as a Kernel, extensible to any FT spectroscopies beyond the ones already implemented.
This approach should help for a \emph{cross-fertilization} in the target spectroscopic domains,
by a wider use of some sophisticated techniques which are currently confined to a specialized domain.

\section{Organisation}
\subsection{Basics about Spike}
SPIKE was developed from the difficulties observed when using the NPK package\cite{Tramesel2007} on large NMR and FT-ICR data-sets (and was even called NPKV2 in the beginning of the project).
It was conceived as a complete rewrite, with a totally new organization while retaining some of its modularity and data concepts.
The choice of the computer language was considered with an emphasis on robustness, flexibility and evolvability, with some consideration on the inherent speed.
Scientific Python\cite{Oliphant2007,Perez2011} was thus chosen as a development platform since it comes with all the benefits of a clear syntax, easy maintainability, simple and fast extensibility.
Processing speed is insured by the scientific modules {Numpy}\cite{Walt2011} and {Scipy}\cite{Jones2001} which have been developed on the top of performant standard libraries developed in low level languages.

All the processing in SPIKE relies on a chained syntax in which each element performs an elementary processing task.
This gives the possibility to write code easily without any programming expertise.
The language can be used for writing scripts or in an interactive
programming context such as the {IPython Notebook}\cite{Perez2007} web interface.
SPIKE comes also with a plugin mechanism which is practical for adding in a flexible manner new functionalities to the native library, without modifying the internal code.

\subsection{Basic Processing}

The elementary basic block routines in SPIKE are apodisation, zero-filling and a full set of Fourier Transforms.
1D and 2D spectra are handled, and the processing of higher dimension spectra is in development.
The handling of hypercomplex numbers originally incorporated for multidimensional NMR
\cite{States1982,delsuc1988} is available and used for the processing of 2D-NMR as well as 2D-FTICR-MS data-sets.
Baseline correction can be done using classical {Savitzky-Golay}\cite{Golay1964} filter or with a specific algorithm using $\ell_1$ norm optimization.
Elementary peak-picking with centroid correction can be performed both on 1D and 2D spectra.

\subsection{Advanced Processing}
Several advanced processing methods are also proposed in the SPIKE program.

It contains specific algorithms such as the Burg algorithm for Linear Prediction\cite{Burg1967,Koehl1999} which can be useful in many situations such as missing points reconstruction or spectral analysis.

Denoising is a major concern in data processing.
Statistical methods such as Cadzow\cite{Cadzow1988} or wavelet denoising\cite{Donoho:1995ue} are very efficient for coping with noise in different kind of signals and are provided here\cite{Brissac:1995we,Agthoven:2011jp}.
SPIKE also incorporates the recently proposed algorithm urQRd\cite{Chiron2014} which
relies on low rank matrix approximation performed with random projections.
This algorithm provides a more robust noise filtering as well as a much faster speed than algorithms of the same kind.
Its much reduced memory footprint permits to handle very large data-sets.
It is used for example on the 2D-FTICR-MS data-set for removing the strong scintillation noise from the 2D-MS data-sets.

\subsection{Data management}
Whenever practical SPIKE performs its data processing in memory.
This is done with the Numpy library which allows vectorized computations of large binary arrays.
In the case of operations too large to fit into memory, the processing is performed \emph{onfile} thanks to the Hierarchical Data Format file format HDF5\cite{hdffive}.
Data stored in this format are accessible transparently as Numpy arrays thanks to the {Pytable}\cite{pytables} Python module.
This feature allows to efficiently cope with the with RAM bottleneck for memory greedy application such as 2D-FTICR-MS processing and visualization.
The HDF5 format was chosen as the default file format for storing SPIKE data.
Its hierarchical structure allows an easy storage of complex data and structured metadata, and the internal compression capabilities insure an optimal file size.

Each manufacturer data-set comes with its own specific importer integrated via a plugin mechanism.
The importers encapsulate the data and metadata in a unique corresponding Python object which
can then be easily manipulated through pipe commands.
The metadata are interpreted from the parameter files provided by the instrument, and give information about characteristic acquisition parameters.
Importers are provided for ThermoFisher \emph{.dat} files, Bruker \emph{Apex} and \emph{Solarix} files, 
as well as \emph{mzXML} generic format; and for Bruker \emph{fid}, \emph{ser} and processed NMR files.
Data can also be easily imported from and exported to the Gifa and NPK software\cite{Pons1996, Tramesel2004} file formats as well as csv text files.

\subsection{Parallelism}
With the emphasis on Big Data processing approach relies on fast algorithms but also on parallel calculations.
This is done transparently using either the multiprocessing Python module for a simple parallelism on multicore machines, or the MPI tool\cite{Gabriel2004} which permits a deployment on a large scale clusters. 

\begin{figure}
\centering
\includegraphics[width=0.4\textwidth]{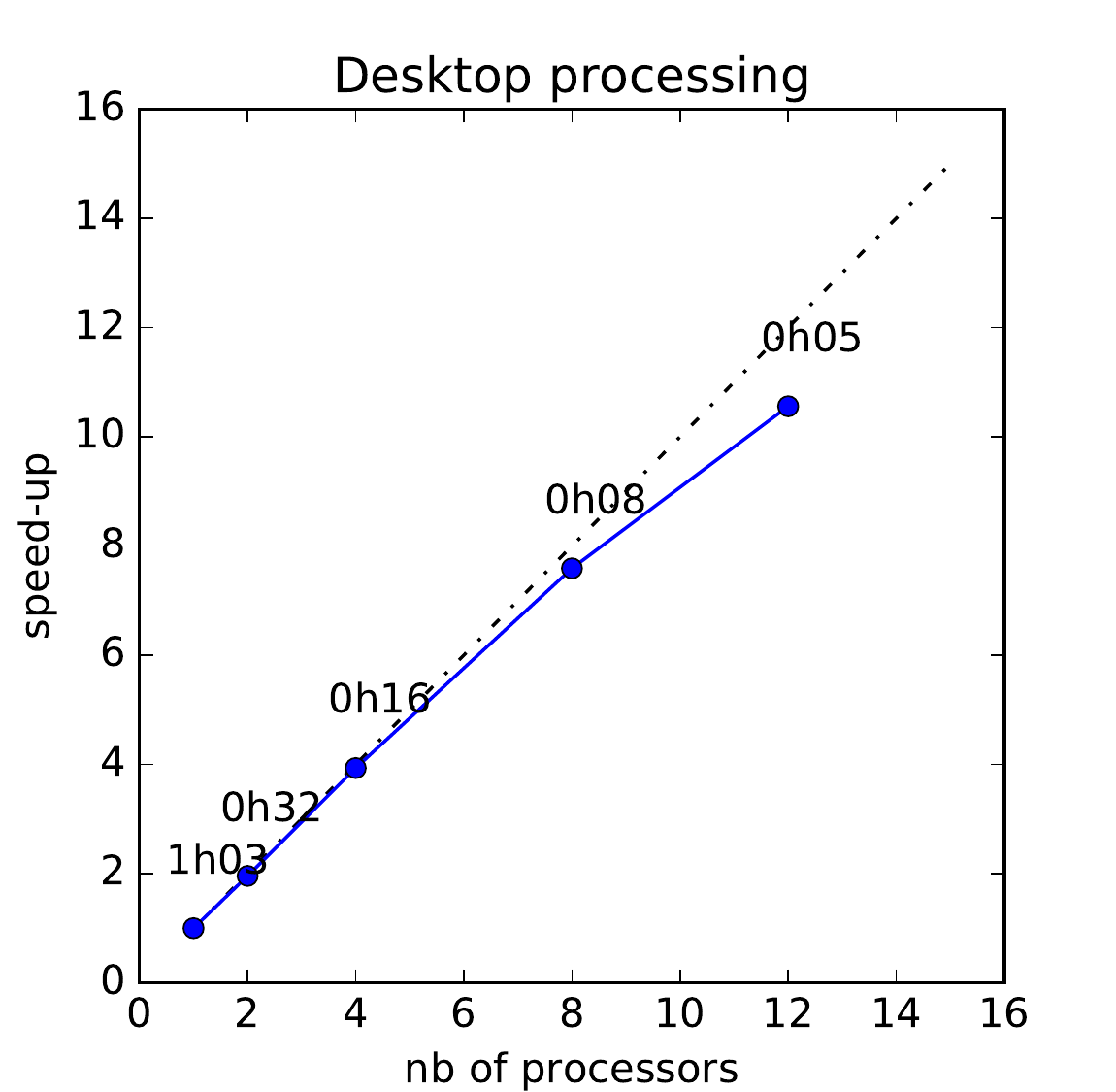}
\includegraphics[width=0.4\textwidth]{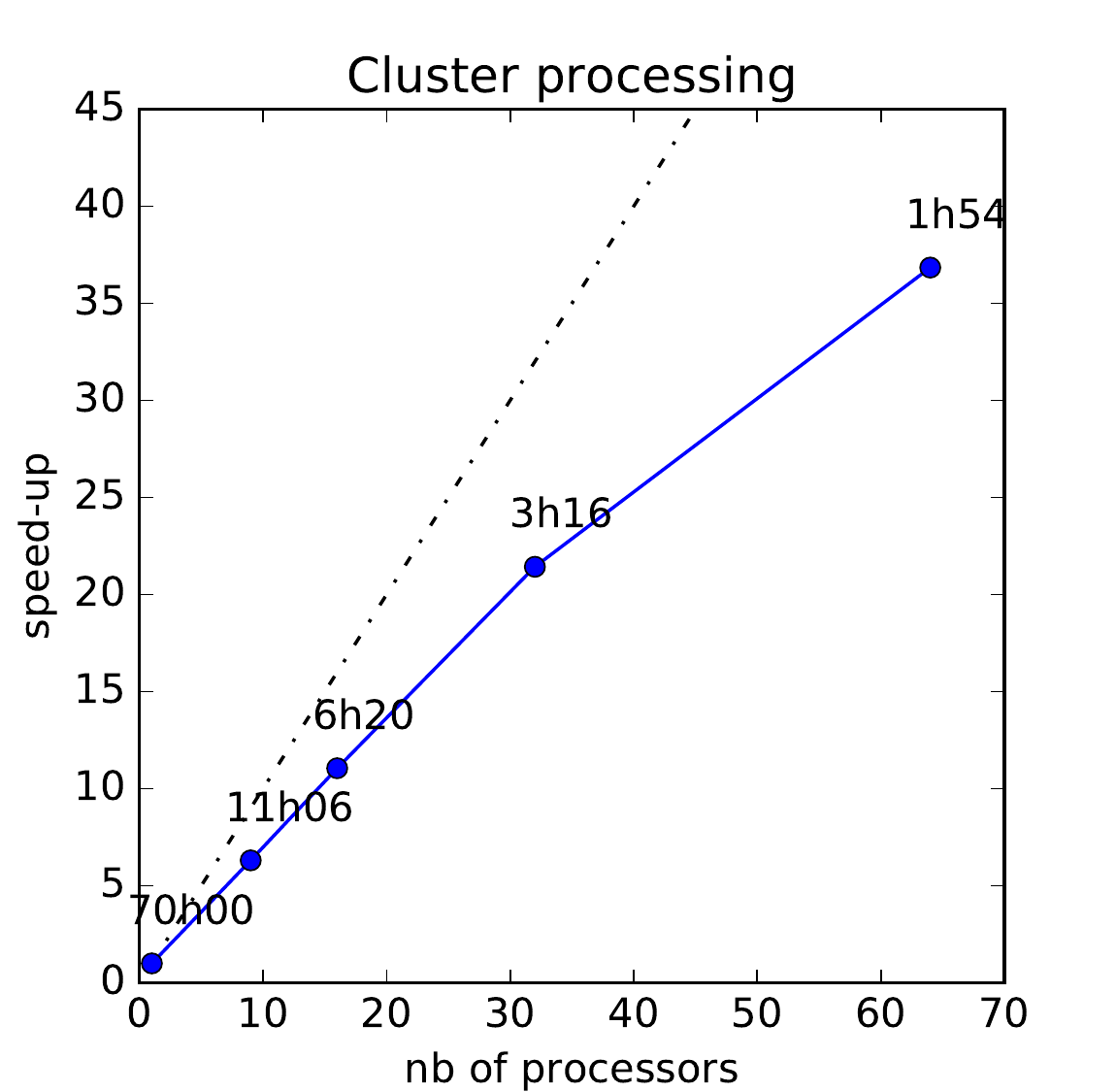}
\caption{Speed-up and processing times for the processing of a 2D-FTICR-MS data-set with SPIKE, the dotted line indicates the $y=x$ line, corresponding to a perfect speed-up.
top) processing of a 2k$\times$64k data-set, (FFT and rank 20 urQRd) to produce a 4k$\times$128k output spectrum,
on a dual Xeon machine with a total of 12 cores, using the python multiprocessing module;
bottom)
processing of a 4k$\times$256k data-set, 
(FFT and rank 200 urQRd) to produce a 8k$\times$512k output spectrum, performed on the Strasbourg University HPC using MPI; the limiting burden here is the file access time.}
\label{fig:parallel}
\end{figure}

A generic interface to MPI was developed inside SPIKE, allowing a very efficient parallelism.
Figure \ref{fig:parallel} presents the processing of a large 2D-FTICR-MS experiment, parallelized on up to 64 processors on a large Linux Beowulf cluster.

\subsection{2D-FTICR-MS}
Recently the equivalent of two-dimensional NMR applied to FTICR-MS, namely 2D-FTICR-MS\cite{Pfand1987} was revived\cite{Agthoven2012,Agthoven2013}.
This is a new exciting but challenging tool for the analysis of very complex mixtures.
In comparison to tandem Mass Spectrometry, no chromatographic separation nor ion isolation are required,  acquisition is greatly simplified and the technique is much more flexible.
Even though it was proposed in the mid eighties, it was not usable in practice until very recently principally because of the limitations on processing and of the huge memory consumption.
Indeed, the latest examples of 2D-FTICR-MS applied to proteomics studies\cite{vanAgthoven:2015gs,Simon:2015by,vanAgthoven:2016kr} would not have been possible without the development of the project presented here.
%This technique became again of interest after many breakthroughs such as the conception of \emph{urQRd}\cite{Chiron2014} for fast denoising and the use of algorithms to increase vertical resolution allowing by the same way to reduce the acquisition time.
%The explosion of the available computational power for desktop computers since the first 2D-FTICR-MS proposal in the mid eighties played also a decisive role for the revival of 2D-FTICR-MS.

A special independent module is provided which performs the complete processing of 2D-FTICR-MS experiment using SPIKE's library.
This module makes an optimal use of all capabilities of the program, in particular the parallel processing.
The processing of this large data-sets is not interactive, so a batch processing mode was chosen and the user has to set-up a configuration file for the parameters for the processing.
This configuration file, as well as the log of the process, is stored as metadata into the final HDF5 file.

\subsection{Visualization}

All data-sets can be displayed interactively using the python matplotlib library\cite{Hunter2007} in a transparent manner, it comes with all the standard tools for manipulating the graph (zoom, drag, saving etc...).
Axes representation is handled according to data type and in their own natural unit (\emph{second}, \emph{Hz}, \emph{ppm}, \emph{m/z}, etc...) 
On multidimensional data the axes can be set independently.

SPIKE proposes also a specific graphic interface developed using the {Qt} library\cite{blanchette2006} dedicated to the display and analysis of the 2F-FTICR-MS spectra (cf Figure \ref{visu2D}).
This module comprises tools for fluid navigation in the large spectral matrix.
Several resolution of the same 2D spectrum are stored in the hierarchical multiresolution HDF5 file structure. 
This allows rapid zooming and the extration of MS subspectra profiles with the highest possible precision without requiring excessive amount of memory despite the huge size of the data matrix.

\section{Examples}
\subsection{Standard 1D spectroscopy}

\begin{figure}
\centering
\includegraphics[width=0.45\textwidth]{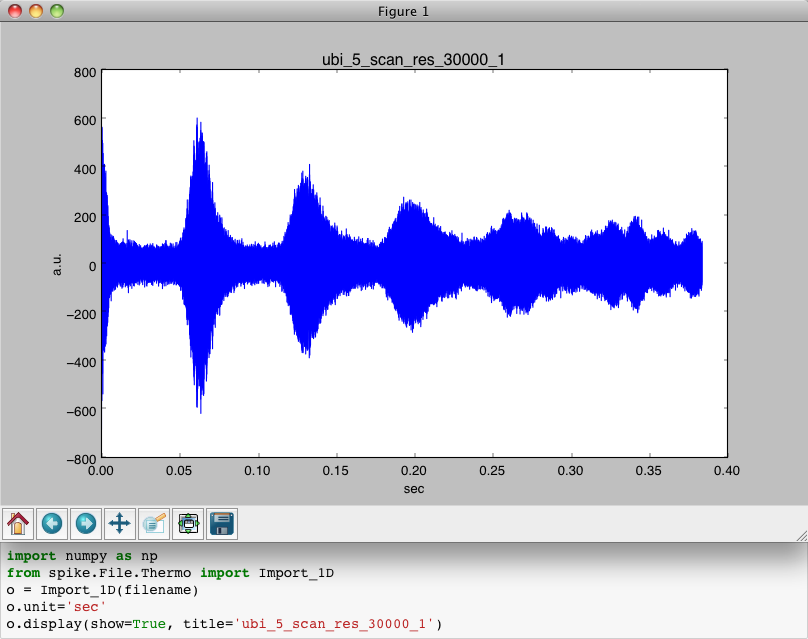}
\includegraphics[width=0.45\textwidth]{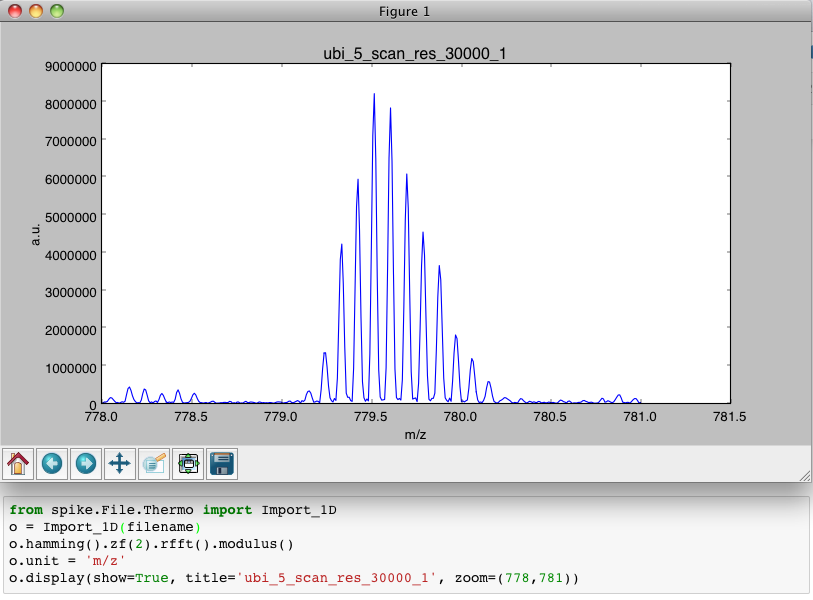}
\caption{Two views of the processing of an Ubiquitin Orbitrap experiment and the corresponding SPIKE commands;
top) importing the data-set and presenting the transient; 
bottom) zoom on the +11 charge state after hamming apodisation, zero-filling, Fourier transform, and modulus.
}
\label{Orbi_fid}
\end{figure}

Using SPIKE it is easy to devise the analysis of a spectroscopic data-set, in few lines of python using the pipe syntax.
As a first example, 
Figure \ref{Orbi_fid} presents the standard processing of an Orbitrap FT-MS data-set,
and the few commands that were issued to produce the spectrum.

\subsection{2D NMR spectroscopy}

\begin{figure*}
\centering
\includegraphics[width=0.8\textwidth]{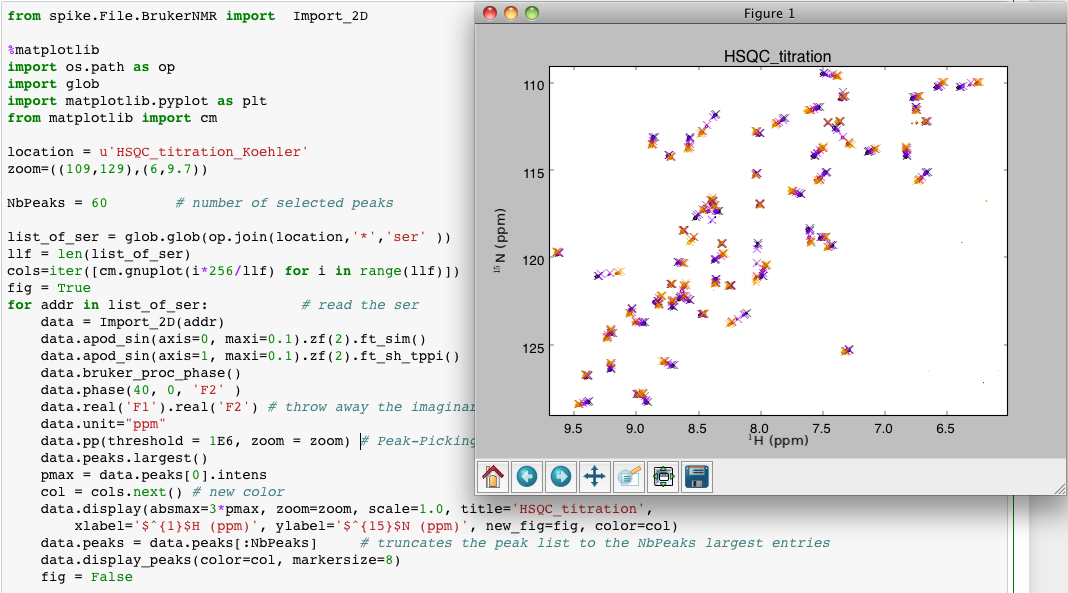}
\caption{Here a series of NMR $^1$H-$^{15}$N HSQC experiments of the interaction of Vinexin$\beta$ protein with a polyproline peptide.
Each raw data-set were imported, processed with apodisation, zero-filling and phase correction,
peak-picked and only the 60 largest peaks were retained,
finally, each spectra and peaks were superimposed on the final picture using a predefined color code.
The code presented here runs in 2-3 seconds on a simple desktop.
}
\label{fig:HSQC}
\end{figure*}

In the NMR spectroscopy studies of protein ligand interactions, a common procedure consists in monitoring the displacement of spectral lines upon some variation of an experimental parameter.
In such an example, it is important to process the spectra in the same condition, to monitor automatically parameters such peak position or width, and finally to report in a graphical manner the evolution of the system.
Figure \ref{fig:HSQC} presents a set of $^{15}$N-HSQC 2D spectra of 
the interaction between the third SH3 domain of Vinexin$\beta$ and varying concentration of a polyproline peptide from the N-terminal domain of the RAR$\gamma$\cite{Koehler:tz}.
The data-sets here are processed and peak-picked automatically.

\subsection{2D-FTICR-MS}

\begin{figure*}
\centering
\includegraphics[width=0.85\textwidth]{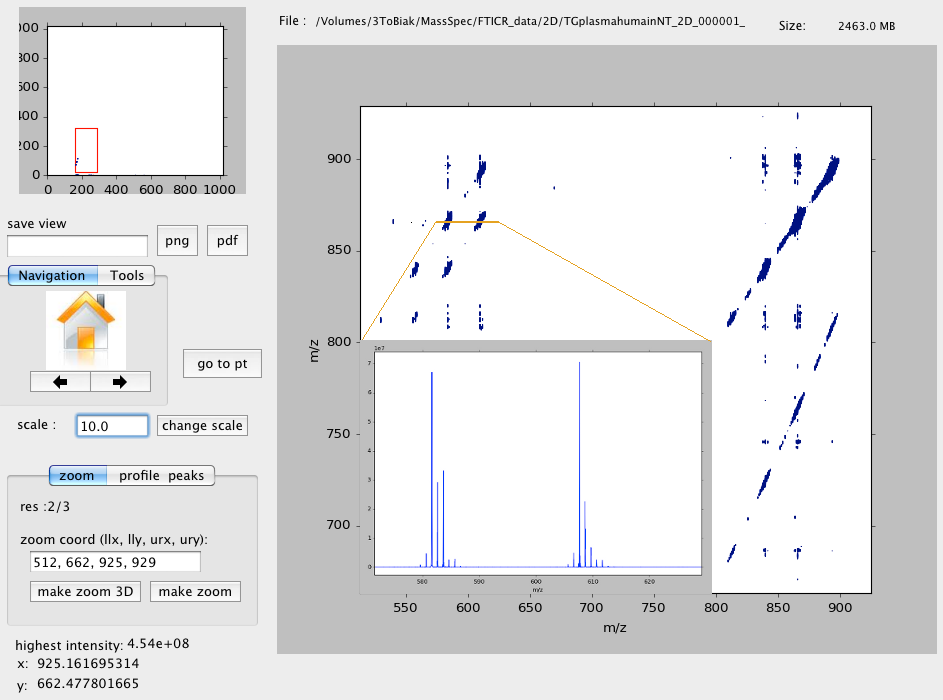}
\caption{Overview of the interface for  2D-FTICR-MS visualization of the 2D IRMPD FT-ICR MS spectrum of triacylglycerols extracted from human plasma;
the middle the main screen presents a zoom on the 2D spectrum, and a pop-up window shows a selected horizontal slice;
the left pane contains various tools (zoom, scale, display history, graphic output, peak-picking, etc.) and information.}
\label{visu2D}
\end{figure*}
In the recent period, the 2D-FTICR-MS method\cite{Pfand1987,Agthoven2012,Agthoven2013}, has 
been shown to be a valuable analytical technique for proteomics studies\cite{Simon:2015by,vanAgthoven:2016kr}.
However, these analyses require the acquisition of very large data-sets, and set high constraint on the analysis software.
SPIKE has been design to handle this kind of large data-sets, and is capable of processing and analyzing  2D-FTICR-MS spectra.

In contrast with the previous examples, the processing is too large to be performed interactively, and specific modules have been developed for the processing and the display of these spectra.

Figure \ref{visu2D} shows the user interface of the program which permits to interactively display 2D-FTICR-MSand extract 1D sub-spectra.
The tool provides rapid zooming and panning, easy extraction of parents and fragments spectra, but also of diagonal spectra (for instance for neutral loss analysis\cite{vanAgthoven:2015gs}).

\section{Conclusion}

We have presented SPIKE, a Python package meant for processing and visualizing data-sets for Fourier spectroscopies with a capacity to easily process large data-sets.
This software aims at giving an easy access to  processing in the domain of FT techniques.
The program is an ongoing development, and
while only NMR and FT Mass Spectrometry are currently available, the code is meant to be extensible to any other FT techniques.

The program SPIKE is entirely written in python, and relies on the standard scientific python libraries.
It is released in open-source under the CeCILL 2.1 license, and available for download at
\url{https://bitbucket.org/delsuc/spike}.

\begin{acknowledgement}

The authors thank C.Köhler and B.Kieffer for the set of NMR HSQC data-sets and J.Chamot-Rooke for the Orbitrap data-set;
the CNRS, for the Défi Instrumention aux limites and the MASTODONS-2013 grant,
the Agence Nationale pour la Recherche (grant FT-ICR2D (2010) and grant  ONE\_SHOT\_FT-ICR\_MS\_2D (2014)) for funding.
The authors acknowledge the High Performance Computing center of the University of Strasbourg for scientific support and access to computing resources.
%Part of the computing resources were funded by the Equipex Equip@Meso project (Programme Investissements d'Avenir).
We also thank all the many colleagues, interns, students, and program users that helped in the development by using the program, and eventually correcting bugs or adding features.

\end{acknowledgement}

% \begin{suppinfo}
% 
% This will usually read something like: ``Experimental procedures and
% characterization data for all new compounds. The class will
% automatically add a sentence pointing to the information on-line:
% 
% \end{suppinfo}

\bibliographystyle{plain}
\small
%\bibliography{spike}
\providecommand*\mcitethebibliography{\thebibliography}
\csname @ifundefined\endcsname{endmcitethebibliography}
  {\let\endmcitethebibliography\endthebibliography}{}

\end{document}